\documentclass[referee]{aa}

\usepackage{graphicx}
\usepackage{amssymb}
\usepackage{natbib}
\usepackage{epsf}

\renewcommand{\ion}[2]{#1\,{\sc #2}}

\newcommand{\lam}{$\lambda$}

\newcommand{\ecs} {erg~cm$^{-2}$~s$^{-1}$} 
\newcommand{\deltE}{\Delta\kern-1ptE}

\title{The Ne/O abundance ratio in the quiet Sun}

\author{P.R.\ Young}

\institute{CCLRC Rutherford Appleton Laboratory, Chilton, Didcot,
  Oxfordshire, OX11 0QX, U.K.}

\date{Received / Accepted}

\abstract
{}
{To determine the neon-to-oxygen abundance in the quiet Sun, a proxy
  for the photospheric abundance ratio.}
{An emission measure method applied to extreme ultraviolet emission
  lines of 
  \ion{Ne}{iv--vi} and 
  \ion{O}{iii--v} ions observed by the Coronal Diagnostic Spectrometer
  on the SOHO satellite.}
{The average Ne/O abundance ratio in supergranule cell centre regions
  is $0.18\pm 0.05$, while in supergranule network regions is $0.16\pm
0.04$. A photospheric Ne/O ratio of $0.17\pm 0.05$ is suggested, in
good agreement with the most recent compilation of solar photospheric
abundances, but discrepant with a recent Ne/O ratio derived
from stellar X-ray spectra and revised neon abundances suggested from
solar interior models.}
{}

\begin{document}

\maketitle

\section{Introduction}\label{sect.intro}

Understanding the physical processes through which the Sun gives rise to
the light that is vital for life on Earth is a fundamental
challenge of astrophysics. The standard models of the Sun's interior
had been considered a great success following the resolution of the
solar neutrino flux problem \citep{ahmad02}, giving excellent agreement with
sound speed and density variation in the solar interior deduced from
helioseismology \citep{bahcall05a}. Recently, however, revisions to the
solar photospheric abundances for the elements carbon, nitrogen,
oxygen  and neon \citep{asplund05} have led to discrepancies between
the models and the observed parameters. Adjustments to the 
solar opacities and element diffusion rates have been ruled out as
solutions to this problem \citep[e.g.,][]{badnell05,guzik05}, and so
attention has focussed on the new 
element abundance values, and in particular the abundance of neon
\citep{antia05,bahcall05b,drake05}.

Unlike other abundant elements, the abundance of neon can not be
determined by analyses of the solar photospheric spectrum as no
absorption lines of Ne or Ne$^+$ are found there. Instead, the neon
abundance has been inferred indirectly from abundance measurements of
solar energetic particles. In such measurements, the neon abundance is
referred to oxygen and it is the downward revision of the oxygen
abundance by 0.17~dex that is largely responsible for the change in
the neon abundance from
$\log\,{\rm Ab}({\rm Ne})=8.08$\footnote{On the scale in which $\log\,{\rm Ab}({\rm H})=12$}  to 7.84 \citep{grevesse98,asplund05}. By
varying parameters in solar 
models, \citet{antia05} have suggested that the models and observations
could be reconciled by increasing the neon
abundance  up to values $\log\,{\rm Ab}({\rm Ne})=8.24$--8.44, and further work by
\citet{bahcall05b} have suggested a value of $\log\,{\rm Ab}({\rm Ne})=8.29\pm
0.05$. 
Independently, \citet{drake05} have measured Ne/O ratios in the
atmospheres of a sample of active stars giving an average value of
0.52, leading to a neon abundance of
$\log\,{\rm Ab}({\rm Ne})=8.27$ (assuming the solar photospheric oxygen
abundance). Following this last work, it is reasonable to ask what is
the Ne/O ratio in the Sun's atmosphere.

The solar atmosphere may not seem a promising place to measure
photospheric abundances
since many years of measurements have demonstrated non-photospheric
abundances in the transition region and corona \citep[see, e.g.,][for
a review]{feldman00}. However, the abundance anomalies are found to
correlate with the first ionization potential (FIP) of the elements,
and neon and oxygen are generally considered as high-FIP elements. In
addition, the average quiet Sun does not show the FIP effect
\citep{young05}, and so derived abundances should reflect the
photospheric values.

The present work uses extreme ultraviolet spectra obtained from quiet
Sun regions by the Coronal
Diagnostic Spectrometer (CDS) on board the SOHO satellite to
determine the Ne/O abundance ratio in the temperature range
$4.6\le\log\,T\le 5.8$ of the Sun's atmosphere. The derived value is
considered a proxy of the photospheric Ne/O ratio.

\section{Data}

The same data-sets were used for this analysis as for that
of \citet{young05}, i.e., 24 sets of SOHO/CDS spectra obtained over a
28 month period from 1996 March to 1998 June. For each observation,
spectra were spatially separated into supergranule network or cell
centre regions based on the intensity of the \ion{O}{v} \lam629.7
emission line. The spectra in each region were then averaged, leading
to 24 sets of network and cell centre spectra. 

\begin{figure}[h]
\includegraphics[scale=0.55]{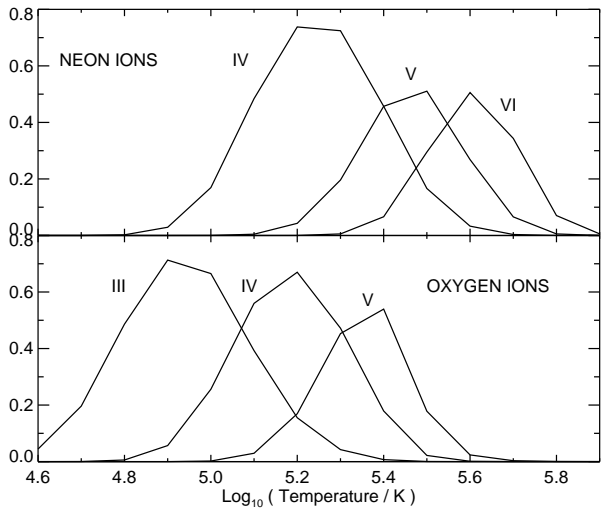}
\caption{Ionization fraction curves from \citet{mazz98} for the
  neon
  and oxygen ions.}
\label{fig.ne-o-ions}
\end{figure}

Temperature overlap between the oxygen and neon ions occurs for
the ions \ion{O}{iii--v} and \ion{Ne}{iv--vi} seen by CDS
(Fig.~\ref{fig.ne-o-ions}). The atomic transitions and wavelengths of
the neon lines were given in \citet{young05}, while those for the
oxygen lines are given in Table~\ref{tbl.transitions}. The emission line
intensities of the neon ions and \ion{O}{v} were previously measured
by \citet{young05}, and so only the \ion{O}{iii} \lam599.6 and the
four \ion{O}{iv} lines were measured here. The four \ion{O}{iv} lines
are partially 
blended with each 
other, and the intensity of the total feature was measured and
treated as a single line in the rest of the analysis. 

The intensities of the oxygen lines and the \ion{Ne}{vi} line were
corrected for the narrow slit burn-in \citep{lang02}, and statistical
errors from the 
line-fitting were added in quadrature to relative uncertainty errors
in the wavelength calibration. The latter vary with
wavelength for the different lines from 20~\% to 29~\% \citep{lang02}.

\begin{table}[h]
\caption{Oxygen emission lines used in the present analysis. Transitions
  within 0.4~\AA\ are blended in the CDS spectra. Wavelengths are from
  v5.1 of the CHIANTI database.}
\begin{flushleft}
\begin{tabular}{lll}
\hline
\hline
\noalign{\smallskip}
Ion &Transition &$\lambda$/\AA  \\
\hline
\noalign{\smallskip}
\ion{O}{iii}&2s$^2$ 2p$^2$ $^1$D$_{2}$ -- 2s 2p$^3$ $^1$D$_{2}$&599.59\\
\noalign{\smallskip}
\ion{O}{iv}&2s$^2$ 2p $^2$P$_{1/2}$ -- 2s 2p$^2$ $^2$P$_{3/2}$&553.33\\
&2s$^2$ 2p $^2$P$_{1/2}$ -- 2s 2p$^2$ $^2$P$_{1/2}$&554.08\\
&2s$^2$ 2p $^2$P$_{3/2}$ -- 2s 2p$^2$ $^2$P$_{3/2}$&554.51\\
&2s$^2$ 2p $^2$P$_{3/2}$ -- 2s 2p$^2$ $^2$P$_{1/2}$&555.26\\
\noalign{\smallskip}
\ion{O}{v}&2s$^2$ $^1$S$_{0}$ -- 2s 2p $^1$P$_{1}$&629.73\\
\hline
\end{tabular}
\end{flushleft}
\label{tbl.transitions}
\end{table}

\section{Method and results}

Following \citet{young05}, the variation of plasma with temperature
through the solar atmosphere 
is modelled through discretising the plasma into isothermal regions
spaced at temperature intervals of 0.1~dex.
The temperature range is chosen to span the regions where the neon and
oxygen ions are formed, i.e., $4.6\le \log\,(T/{\rm K}) \le 5.8$ (see
also Fig.~\ref{fig.ne-o-ions}).
The plasma column depths, $h_i$,
are allowed to vary at only three temperatures: $\log\,(T/{\rm
  K})=4.6, 5.2$ and 
5.8. I refer to these values as $h_0$, $h_6$ and $h_{12}$. Values of
$h_i$ at intermediate temperatures are derived by 
linear interpolation in the $\log\,T$--$\log\,h$ plane during the
minimization procedure. In deriving the $h_i$ values, the absolute
abundance of one element must be assumed, and I take the new
photospheric abundance of oxygen: [O/H]=8.66 \citep{asplund05}.

\begin{figure}[h]
\includegraphics[scale=0.55]{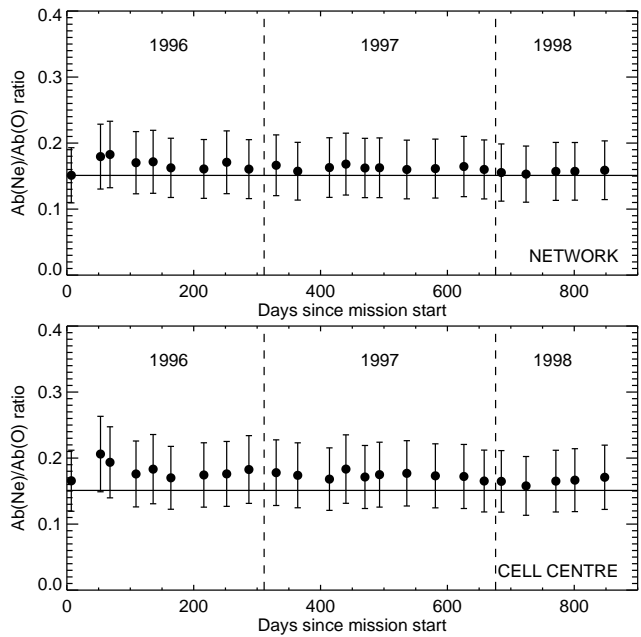}
\caption{The derived Ne/O abundance ratios for the network (upper
  panel) and cell
  centre (lower panel) regions as a function of time. The black
  horiontal line denotes the photospheric Ne/O abundance ratio
  \citep{asplund05}.}
\label{fig.neo-ratio}
\end{figure}

There are thus four parameters -- Ab(Ne)/Ab(O), $h_0$, $h_6$ and $h_{12}$
-- to be fit to the six observed oxygen and neon intensities. The line
intensities are modelled through Eq.~1 of \citet{young05}, and atomic
data are from v.5.1 of the CHIANTI database \citep{landi05}. The 
minimization procedure of \citet{young05} is applied, and the Ne/O
abundance ratios for each of the 24 data-sets in the network and cell
centre regions are displayed in Fig.~\ref{fig.neo-ratio}.

The statistical average of the Ne/O abundance ratio in the cell centre
regions is $0.18\pm 0.01$, while in the network regions it is
$0.16\pm 0.01$. The derived values  over the 28 month period are
remarkably consistent.

A fixed pressure of $10^{14.5}$~K cm$^{-3}$ was assumed for the
analysis \citep[see][]{young05}, but assuming a constant density
instead does not make a significant difference to the results, as
shown in Table~\ref{tbl.var} where results for densities of $10^9$ and
$10^{10}$~cm$^{-3}$ are shown.

Atomic data uncertainties have not been included in
the analysis and are likely to be significant, particularly for the
ionization fractions of the ions. To investigate the effects of
modified ion fractions in a simplistic manner, the analysis was
repeated by shifting the ion 
fractions, $F$, of only the neon ions forwards and backwards in temperature
by 0.1~dex. I.e., $F^\prime(T_i)=F(T_{i-1})$ or
$F^\prime(T_i)=F(T_{i+1})$, respectively. The results of this are
shown in Table~\ref{tbl.var}, where $F_{\rm Ne}+$ indicates the neon
ions have been moved forward in temperature, and $F_{\rm Ne}-$
backwards in temperature. The effects are again small, and we use
these results to give final error bars of $0.18\pm 0.05$ and $0.16\pm
0.04$ on the Ne/O relative abundance in cell centre and network regions.

To demonstrate consistency with the analysis of \citet{young05}, the
$h_i$ values from a single data-set are compared with those from
Fig.~3 of \citet{young05} and excellent agreement is found in the
overlap region. The column depths from the Ne/O analysis have been
scaled downwards by a factor 0.263 due to the different reference
abundance used \citep[oxygen in the present analysis, and neon in
the][analysis]{young05}.

\begin{figure}[h]
\includegraphics[scale=0.55]{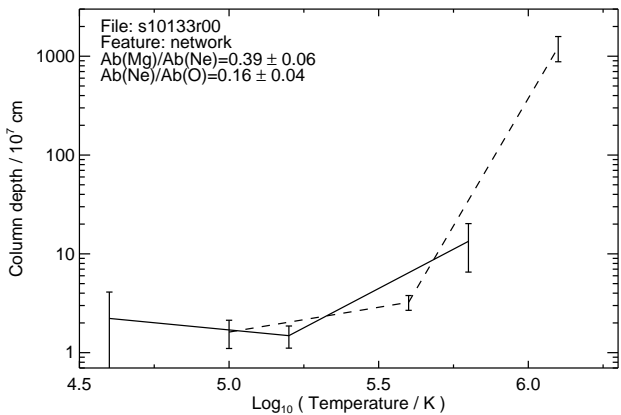}
\caption{A comparsion of column depths ($h$) derived from the present
  analysis (solid line) with those from \citet{young05} (dashed line) for the
  network spectrum 
  from CDS data-set s10133r00. The curves are scaled to remove the
  dependence on the reference abundances (see text).}
\label{fig.emd-figure}
\end{figure}

As a further check on the reliability of the results, the derived
column depths were used to predict the intensity of the \ion{O}{vi}
\lam1032 line in cell centre and network regions using atomic data
from CHIANTI. Although not
observed by CDS, this line was measured by the Harvard S055 instrument
on board \emph{Skylab} and \citet{vernazza78} give average intensities
of 474 and 223~\ecs\ in network and cell centre regions,
respectively. Averaging the predicted intensities from each of the 24
data-sets here gives values of $487\pm 60$ and $211\pm 26$, in
excellent agreement with the \citet{vernazza78} values.

\begin{table}[h]
\caption{Ab(Ne)/Ab(O) values derived for different assumptions: fixed
  pressure, fixed density, and displacements of the neon ion fractions.}
\begin{flushleft}
\begin{tabular}{lll}
\hline
\hline
\noalign{\smallskip}
 &Cell centres &Network \\
\hline
\noalign{\smallskip}
$P=10^{14.5}$~K cm$^{-3}$ &$0.18\pm 0.01$ &$0.16\pm 0.01$ \\
$N_{\rm e}=10^{9}$~cm$^{-3}$ & $0.19\pm 0.01$& $0.18\pm 0.01$\\
$N_{\rm e}=10^{10}$~cm$^{-3}$ & $0.19\pm 0.01$& $0.18\pm 0.01$\\
$F_{\rm Ne}+$ &$0.13\pm 0.01$ & $0.12\pm 0.01$ \\
$F_{\rm Ne}-$ &$0.22\pm 0.01$ & $0.20\pm 0.01$ \\
\hline
\end{tabular}
\end{flushleft}
\label{tbl.var}
\end{table}

\section{Discussion}

Averaging the derived abundance ratios for the supergranule network
and cell 
centre regions gives a value of Ab(Ne)/Ab(O) of $0.17\pm 0.05$, in
excellent agreement with the ratio of $0.15\pm 0.03$
obtained from the solar photospheric abundance tables of
\citet{asplund05}. No evidence 
is found for the enhanced Ne/O abundances found by \citet{drake05}
from analyses of X-ray spectra of active cool star atmospheres.

The advantage of studying the quiet Sun is that it is relatively
stable compared with the hot active region (likely flaring) plasma
that gives rise to the X-ray neon and oxygen lines in active
stars. This is reflected in the remarkably consistent value of the
Ne/O ratio over the 28 months of CDS observations
(Fig.~\ref{fig.neo-ratio}). Observations of solar flares
have  
demonstrated Ne/O variations of a factor 2 between different events
\citep[e.g.,][]{fludra95}, 
while stellar 
atmospheres show abundance patterns not consistent with the solar
corona \citep[e.g.,][]{drake01}, which could imply as yet unknown
processes causing the modifications from the solar photospheric
Ne/O ratio.

Using the photospheric abundance of oxygen from \citet{asplund05}
my results lead to a [Ne/H] value of $7.89\pm 0.16$, a value not
consistent 
with the neon abundances suggested by \citet{antia05} and
\citet{bahcall05b}.

The conclusion from the present analysis is thus that the photospheric
abundance of neon is \emph{not} responsible for the discrepancies
between standard solar models and observations.

\end{document}